\pdfoutput=1 
\documentclass{nature}
\usepackage{lineno}
\usepackage{underscore}
\usepackage{multirow}
\usepackage{indentfirst}

\usepackage{ccaption}
\usepackage{subcaption}
\usepackage{graphicx}
\usepackage{lipsum}
\usepackage{ragged2e}
\usepackage{hyperref}

\usepackage{cmbright}
\usepackage{filecontents}
\usepackage{amsmath}

\newcounter{lastnote}

\usepackage{comment}
\usepackage{cite}
\usepackage{adjustbox} 

\makeatletter
\newenvironment{figurehere}
{\def\@captype{figure}}
{}
\makeatletter

\usepackage{xcolor}
\definecolor{sared}{rgb}{0.69, 0,0}

\title{\begin{center}  Vertical NAND in a Ferroelectric-driven Paradigm Shift \end{center}}

\author
{ \centering
{
\large{Giuk Kim$^{1,2}$, Hyojun Choi$^{2}$, Prasanna Venkat Ravindran$^{3}$, \\ Moonyoung Jung$^{4}$, Sanghyun Park$^{4}$, Kijoon Kim$^{4}$, Suhwan Lim$^{4}$, Kwangyou Seo$^{4}$, Kwangsoo Kim$^{4}$, Wanki Kim$^{4}$, Daewon Ha$^{4}$, Sukjoong Shin$^{2}$,\\Asif Khan$^{3}$, Sanghun Jeon$^{2\dag}$, Kai Ni$^{1\dag}$\\}
\vspace{3ex}
\normalsize{$^{1}$University of Notre Dame, Notre Dame, IN 46556, USA}\\
\normalsize{$^{2}$Korea Advanced Institute of Science and Technology, Daejeon, 34141, Korea;}\\
\normalsize{$^{3}$Georgia Institute of Technology, Atlanta, GA 30332, USA;}\\
\normalsize{$^{4}$Samsung Electronics Co., Ltd,
Hwaseong, 18367, Korea;}\\
\vspace{2ex}
\normalsize{$^{\dag}$To whom correspondence should be addressed} \\
\normalsize{Email: jeonsh@kaist.ac.kr, kni@nd.edu} \\
}}

\begin{document}
\flushbottom
\maketitle
\vspace{3ex}
\begin{abstract}

Over the past decades, the relentless scaling and mass production of flash memory have underpinned the data-centric era. Yet charge-trap-based 3D NAND flash is now constrained by intrinsic physical and architectural limits, including reliability degradation at the device level, high operating power at the array level, and vertical scaling saturation at the system level. These bottlenecks hinder further advances in storage density and energy efficiency required by memory-centric computing. This Perspective outlines how coupling ferroelectric polarization with charge trapping can reconfigure the foundations of flash memory. In these hybrid architectures, polarization offers an energy-efficient pathway for charge modulation through enhanced Fowler-Nordheim tunneling, while trapped charges reinforce polarization-driven states to ensure stability. Such synergistic dynamics enable low-voltage operation and integration beyond one thousand layers without compromising process compatibility. We discuss the material, device, and architectural transitions required to realize this hybrid technology and chart future research directions to overcome the remaining scaling bottlenecks. Hybrid ferroelectric NAND extends conventional flash toward a scalable and energy-efficient platform, marking a paradigm shift for next-generation non-volatile memory.
\end{abstract}

%
%


\section*{\textcolor{sared}{\large Introduction}}

Over the past four decades, flash memory has evolved into one of the foundational pillars of modern electronics, driven by continuous advances in materials, device, architecture, and manufacturing\cite{lee2007layer, baeg2014charge}. As the dominant non-volatile storage technology in the computing hierarchy, charge-trap-based NAND flash forms the backbone of the data-centric world\cite{liu2021ultrafast, compagnoni2017reviewing, sanvido2008nand, yoon2018nanophotonic, kim2023review, kim2016simultaneous}. Since its commercialization in the 1980s, NAND flash has sustained an extraordinary scaling trajectory, achieving an average annual bit-density growth of about 30 percent\cite{kim2017evolution, krishnan2025nand, goda2024nand}. The transition from planar two-dimensional (2D) cells to vertically stacked three-dimensional (3D) architectures replaced lateral miniaturization with vertical expansion, enabling more than six orders of magnitude increase in storage density and establishing NAND flash as the mainstream platform for non-volatile memory\cite{parat2018scaling, alsmeier2020past, lee2016new, kim2009novel}. Fig. \ref{fig:overview} traces this evolution, highlighting the key milestones from floating-gate to charge-trap structures and ultimately to 3D vertical integration, which together have upheld the scaling trend for decades\cite{aritome2016nand}. From smartphones and solid-state drives to large-scale data centers, NAND flash powers both consumer electronics and cloud infrastructures, serving as a cornerstone of the digital economy and the semiconductor industry. Alongside these technological achievements, NAND flash has also begun to transcend its traditional role as a passive storage medium, increasingly functioning as an active computing element within emerging in-memory architectures tailored for AI-centric systems\cite{hu2022512gb, lee2024review, kim2021embedded, lue2019optimal}

\begin{figurehere}
   \centering
    \includegraphics[width=1\linewidth]{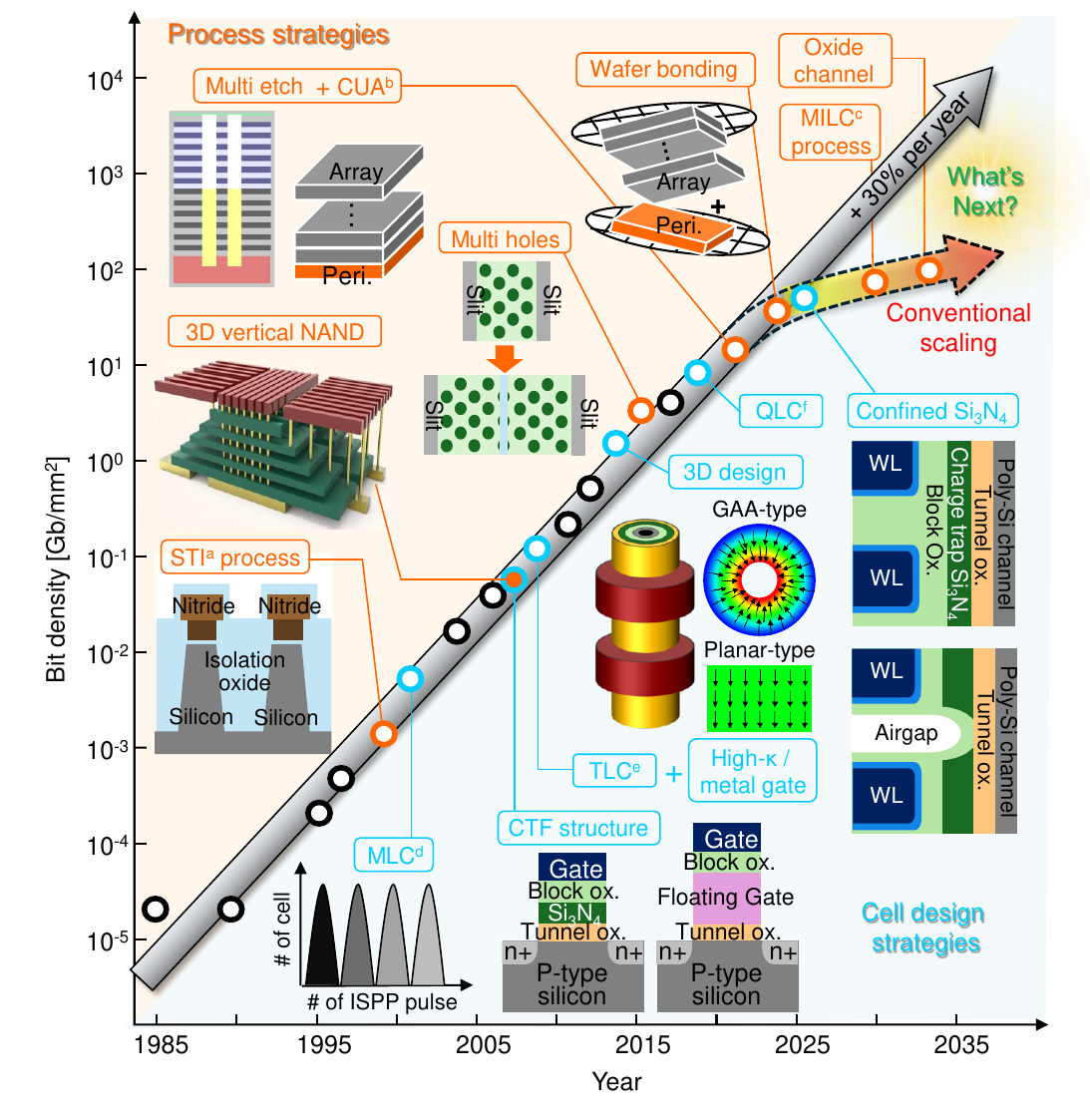}
    \captionsetup{parbox=none} 
    \caption{\textit{\textbf{Historical and future trends of NAND flash bit density.}   Evolution of NAND flash bit density from planar 2D scaling to 3D vertical stacking. Until 2013, conventional planar scaling was pursued down to approximately 15 nm, enabled by shallow trench isolation, multi-level cell operation, and Si\textsubscript{3}N\textsubscript{4} charge-trap technologies. The introduction of 3D vertically stacked NAND flash in 2007 marked a paradigm shift, initiating intensive efforts to integrate more cells within a limited footprint. Manufacturable innovations such as multi-hole processes for increasing string density, multi-etch schemes for stacking multiple strings, and cell-under-periphery (CuA) architectures for reducing peripheral area have since been adopted in mass production. Looking ahead, advances in materials, process technologies, and cell architectures are expected to usher in an era of ultra-high-density storage memories with more than 1,000 vertical layers and bit densities exceeding 100 GB/mm\textsuperscript{2}.  STI\textsuperscript{ a}, shallow trench isolation; CUA\textsuperscript{ b}, cell under array; MILC\textsuperscript{ c}, metal-induced lateral crystallization; MLC\textsuperscript{ d}, multi-level-cell; TLC \textsuperscript{ e}, triple-level-cell; QLC\textsuperscript{ f}, quad-level-cell;}}
    \label{fig:overview}
\end{figurehere}

Despite its remarkable progress, 3D NAND flash has entered an era in which the very mechanisms that once enabled its success now constrain its evolution\cite{lim2023comprehensive, han2023fundamental}. The charge trapping and high-field tunneling processes fundamental to NAND operation can no longer ensure stable and reliable performance. In charge-trap architectures, information is stored as charges trapped within the silicon nitride (Si\textsubscript{3}N\textsubscript{4}) layer, but the high program (PGM) and erase (ERS) voltages required for charge injection inevitably accelerate severe reliability degradation\cite{jo2024investigation, liu2009reliability}. These device-level constraints cascade upward to the array and architectural levels. Attempts to further increase bit density by reducing the pitch of word lines (WLs) have intensified dielectric breakdown in inter-layer space oxides and cell-to-cell interference, rendering additional vertical stacking increasingly impractical\cite{lim2023comprehensive, han2023fundamental, kim2024depth, kim2024unveiling}. Moreover, the large operating voltages induce significant power consumption, reducing the overall energy efficiency of NAND systems\cite{khan2020future,tanzawa2002circuit}. Although modern 3D NAND already integrates three hundreds of stacked layers, each additional layer now amplifies the cumulative physical and electrical limitations inherent to the structure\cite{goda2024nand, lee2023novel, kim202328}. Simply "stacking higher" can no longer sustain the historical trajectory of NAND scaling, underscoring the need for a fundamental reconfiguration of the charge-trap mechanism itself.

In this regard, hafnia-based ferroelectrics have emerged as a compelling pathway to redefine the operational paradigm of 3D NAND technology. These materials exhibit robust, non-volatile polarization switching even at thicknesses below ten nanometers, while remaining fully compatible with both advanced CMOS fabrication and the mature 3D NAND process\cite{schroeder2022fundamentals, kim2021cmos, florent2017first, mukherjee2024improved}. Their intrinsic properties enable the development of ferroelectric transistors and capacitors that store information through polarization rather than trapped charges, thereby addressing the fundamental reliability and scalability challenges inherent to conventional charge-trap flash\cite{fernandes2024material, kim2024exploring}. More importantly, ferroelectric polarization introduces a fundamentally different mechanism from conventional tunneling-based charge trapping, characterized by its strong nonlinearity and high efficiency in field-induced charge modulation.\cite{kim2024depth, kim2024unveiling, qin2024clarifying, qin2025elucidating}. This unique property not only enhances charge injection and storage but also enables cooperative operation with the trapping layer, where the stored charges stabilize the ferroelectric states. Such a synergistic ferroelectric–charge-trap interaction provides a promising pathway to overcome intrinsic efficiency limitations and open a new paradigm of NAND scaling.

This perspective moves beyond the simple insertion of a ferroelectric layer into the gate stack and instead proposes a comprehensive reconfiguration of 3D NAND technology. We explore how hybrid ferroelectric NAND (FeNAND) can serve as a paradigm-shifting platform for next-generation non-volatile memories. Specifically, we compare the operational principles of charge-trap flash, ferroelectric transistors, and hybrid FeNAND devices to demonstrate that their dual-operation mechanism enables concurrent improvements in performance, reliability, and bit density beyond the limits of existing 3D NAND architectures. In addition, we highlight new opportunities spanning gate-stack engineering, array operation schemes, 3D architecture, and the coupling with oxide-semiconductor channels. These aspects, often examined in isolation, are revisited here from an integrated system perspective that unites material physics, device physics, array operation, and architectural implementation, thereby establishing coherent design guidelines for hybrid FeNAND technologies. Finally, this perspective identifies the key performance and process challenges that currently hinder the commercialization of oxide-semiconductor-based hybrid FeNAND and outlines future research directions to address them. Rather than summarizing existing progress, our goal is to chart a forward-looking research and industrial roadmap through which FeNAND can evolve into a central technology driving the next era of electronic innovation.

\section*{\textcolor{sared}{\large Scaling Limitations of 3D NAND Technology}}
 
Fig. \ref{fig:scaling} summarizes the physical origins of the scaling bottlenecks in today’s 3D vertical NAND flash and illustrates the intrinsic trade-off between integration density and reliability. To sustain the historical trajectory of bit-density scaling, hafnia-based ferroelectric materials offer a promising pathway to overcome these limitations. The conceptual framework of hybrid FeNAND architectures, which integrate ferroelectric polarization and charge trapping mechanisms, is presented here as a design map for achieving both scalable and reliable data storage in future memory technologies.

Fig. \ref{fig:scaling}(a) schematically illustrates the key scaling strategy that has enabled the continued increase in bit density of charge-trap-based 3D NAND flash. The conventional architecture adopts a gate stack composed of a block oxide, Si\textsubscript{3}N\textsubscript{4} charge-trap layer, tunnel oxide, and channel. In 3D NAND technology, increasing the string height has become infeasible due to constraints from subsequent packaging processes. To further improve integration density, a scaling strategy has been adopted to gradually reduce the pitch between WLs and the intervening oxide, with the combined height of the WL and space oxide reaching approximately 40 nm \cite{han2023fundamental}. However, this relentless pursuit of vertical scaling introduces fundamental limits to device operation. In charge-trap 3D NAND, data are stored by injecting and retaining charges in the Si\textsubscript{3}N\textsubscript{4} charge-trap layer through Fowler–Nordheim (FN) tunneling under a high electric field. The memory state is determined by the quantity of trapped charge, and achieving sufficient charge injection requires program voltages typically above 20 V\cite{kim2024depth, kim2023opportunity, shin2020capacitance}. Such high-field operation is intrinsic to the charge-trap mechanism itself. As the inter-WL dielectric becomes thinner, its breakdown voltage (V\textsubscript{BD}) inevitably decreases, and the intensified electric field during programming accelerates space-oxide degradation\cite{han2023fundamental, kim2024unveiling, choi2025opportunity}. The resulting premature breakdown undermines device reliability and can ultimately cause functional failure of the entire 3D NAND stack.

\begin{figurehere}
   \centering
    \includegraphics[scale=0.1,width=\textwidth]{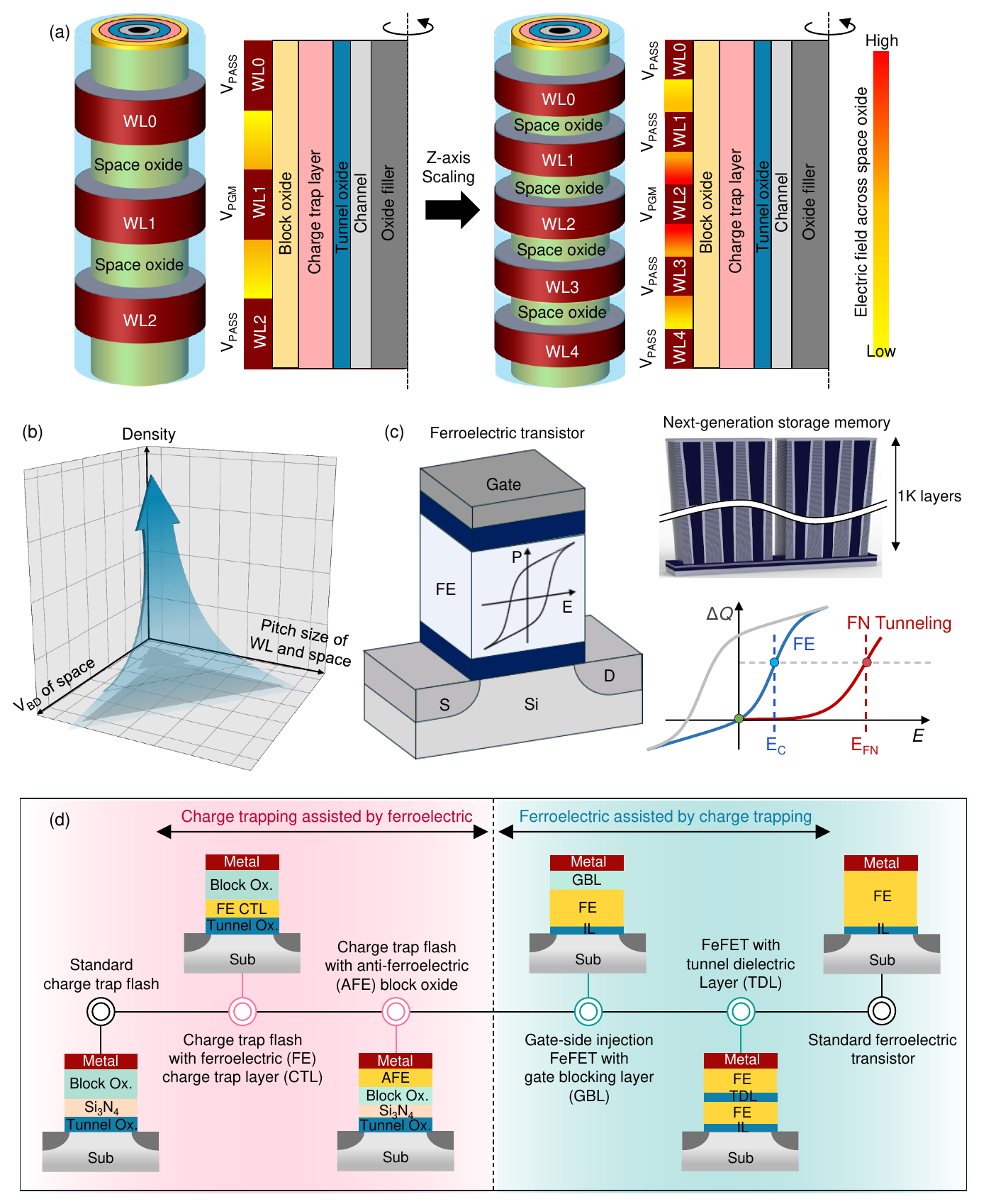}
    \captionsetup{parbox=none} 
    \caption{\textit{\textbf{Overcoming NAND flash scaling limits through hybrid integration.} (a) Scaling strategy of conventional 3D NAND flash, in which minimizing the WL and space-oxide pitch enhances bit density but increases the electric field across the thinned space oxide, lowering its V\textsubscript{BD} and causing PGM-induced dielectric failure within the string. (b) Trade-off between vertical scaling and reliability, showing that reducing the WL and space oxide pitch to boost bit density inevitably degrades the V\textsubscript{BD}, leading to PGM-induced failure. (c) Ferroelectric transistor utilizing polarization-switching mechanisms, offering low-power, high-speed, and energy-efficient operation. These characteristics make it a promising building block for next-generation Ferroelectric NAND architectures and in-memory computing applications. (d) Representative gate-stack hierarchy bridging conventional charge-trap flash and ferroelectric transistors, illustrating FeNAND designs that co-utilize ferroelectric polarization and charge trapping, ranging from charge-trap-dominant to ferroelectric-dominant operation.}}
    \label{fig:scaling}
\end{figurehere}

Fig. \ref{fig:scaling}(b) visualizes the fundamental trade-off between vertical scaling and reliability in charge-trap-based 3D NAND flash. As cell stacking increases and the vertical pitch is reduced, the intensified electric field across the thinner inter-WL dielectric lowers its V\textsubscript{BD}, setting a practical limit on further integration. This relationship reveals that the physical bottleneck of NAND scaling arises from the high-field FN tunneling mechanism that underlies charge-trap operation. As a result, continued vertical stacking within the conventional NAND framework becomes increasingly impractical, underscoring the need for a low-voltage storage paradigm that transcends the limits of charge trapping.

Responding to the need for a low-voltage storage paradigm, hafnia-based ferroelectrics have emerged as a promising route to overcome the scaling limitations of 3D NAND flash. Ferroelectric transistors store information through non-volatile polarization switching, where controlled partial switching enables multi-level memory operation\cite{si2019ferroelectric, kim2023ferroelectric, mulaosmanovic2018mimicking, ni2019ferroelectric}. Ferroelectric NAND flash, which integrates polarization functionality into a charge-trap architecture, represents a new class of non-volatile memory. Here, dynamic polarization switching not only encodes information, but also facilitates low-voltage charge tunneling, markedly reducing the operating voltages. Thanks to these characteristics, conventional charge-trap-based NAND suffers from a physical limitation that prevents the incremental step pulse programming (ISPP) slope from exceeding 1, whereas ferroelectric NAND flash achieves a high ISPP slope above 2, demonstrating superior program efficiency\cite{kim2024depth, kim2024unveiling, kim2023opportunity, shin2020capacitance}. Furthermore, hafnia ferroelectrics provide excellent thermal stability, precise thickness control via atomic layer deposition (ALD), and full compatibility with standard CMOS processing, making them well suited for highly integrated 3D fabrication\cite{deng2023comparative, deng2024first, chang2020anti}. Collectively, these advantages position hafnia ferroelectrics as a key enabler for extending 3D NAND technology toward beyond 1000-layer vertical stacking\cite{lee2024analog, kim2023highly, song2025ferroelectric} (Fig. \ref{fig:scaling}c). Fig. \ref{fig:scaling}(d) depicts the conceptual classification of hybrid FeNANDs, which bridge the operational characteristics of conventional charge-trap flash and ferroelectric transistors. These hybrid FeNAND architectures can be broadly divided into charge-trap-dominant and ferroelectric-dominant types, where polarization and charge trapping respectively play complementary roles. Looking ahead, research efforts should focus on developing design strategies that simultaneously optimize reliability, operating characteristics, and 3D architecture, guided by a fundamental understanding of the dynamic coupling between polarization switching and charge trapping.

\section*{\textcolor{sared}{\large Shaping the memory landscape with hybrid FeNAND}}

Building on the conceptual classification, this section provides an in-depth analysis of four representative flash memory devices distinguished by their charge-storage and polarization-switching mechanisms. Conventional charge-trap flash stores information purely through carrier trapping in a Si\textsubscript{3}N\textsubscript{4} dielectric, whereas ferroelectric flash relies on polarization reversal within hafnia-based ferroelectrics. The combination of these two distinct mechanisms gives rise to hybrid FeNAND architectures, which can be broadly divided into two categories: those in which ferroelectric switching assists charge trapping, and those in which charge trapping assists ferroelectric polarization. Understanding the dynamic interplay between polarization switching and charge trapping in such hybrid systems is critical for optimizing device performance, array operation, and 3D integration. Fig. \ref{fig:fenand} compares the gate-stack configurations, operation principles, and analytical models of these four device types, highlighting the intrinsic interdependence between memory window (MW) and electrical operation characteristics.

The standard charge-trap flash employs an oxide–nitride–oxide (ONO) gate stack and operates solely through charge trapping (Fig. \ref{fig:fenand}a).
During PGM operation, a high electric field across the tunnel oxide induces FN tunneling, injecting electrons from the channel into the Si\textsubscript{3}N\textsubscript{4} charge-trap layer and thereby forming a high-threshold-voltage (HVT) state (Fig. \ref{fig:fenand}b). ERS operation proceeds through hole injection into the same layer, establishing a low-threshold-voltage (LVT) state (Fig. \ref{fig:fenand}c). Both processes rely on FN tunneling, whereas READ and PASS operations occur under lower biases, during which tunneling is suppressed and the trapped charge ideally remains constant. Fig. \ref{fig:fenand}d presents the band diagram under 

\begin{figurehere}
   \centering
    \includegraphics[scale=0.1,width=\textwidth]{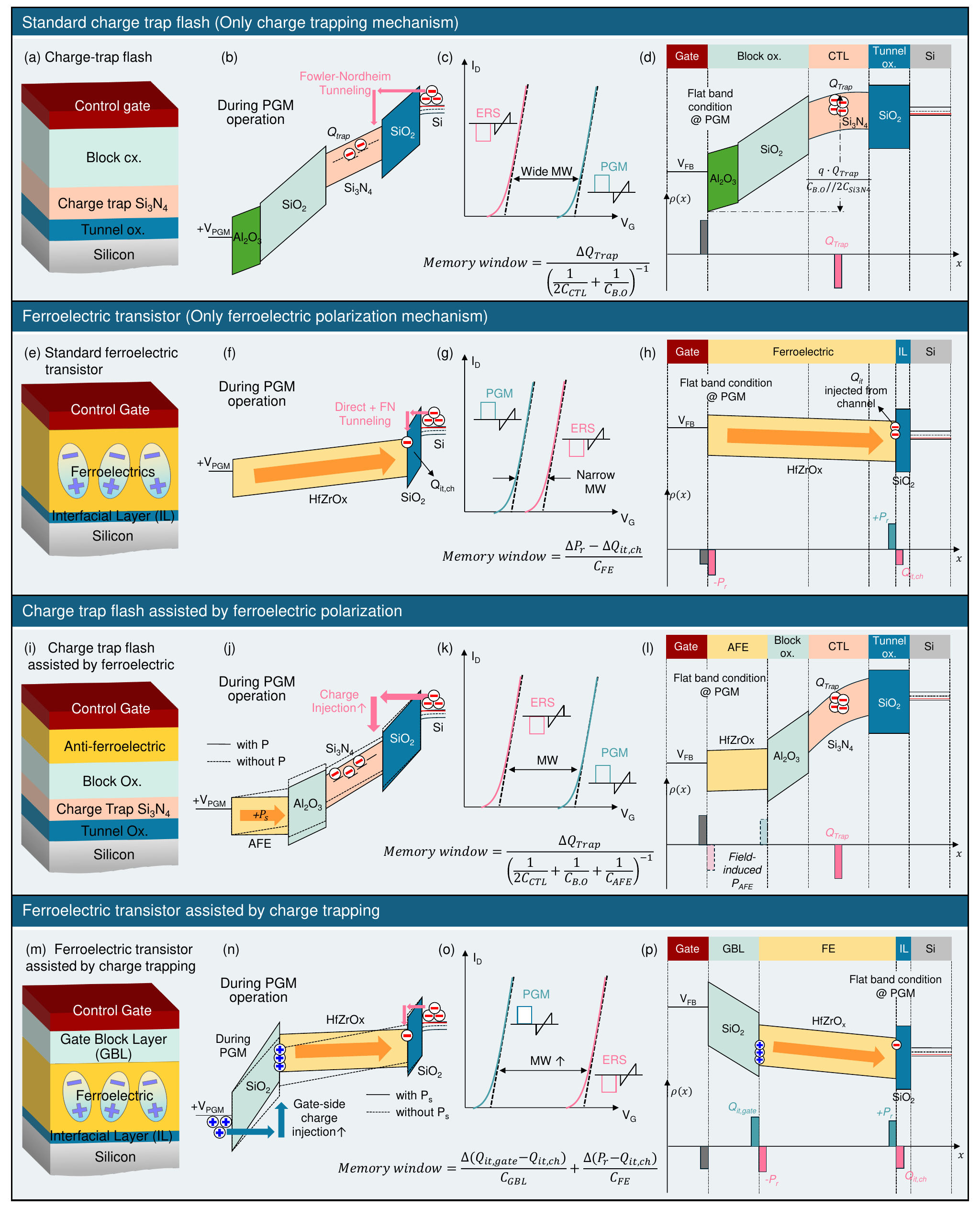}
    \captionsetup{parbox=none} 
    \caption{\textit{\textbf{Gate-stack hierarchy and mechanisms of charge-trap, ferroelectric, and hybrid memories.} (a-d) Standard charge-trap flash. Gate stack and PGM mechanism based on channel-side charge injection and trapping. V\textsubscript{TH} shifts after PGM and ERS define the HVT and LVT states that constitute the MW. Flat-band energy diagram illustrating trapped charge effects and the corresponding analytical model. (e-h) Standard ferroelectric transistor. Gate stack and polarization-switching-based PGM operation. V\textsubscript{TH} shifts after PGM and ERS exhibit inverted polarity (LVT/HVT) compared with charge-trap flash. Flat-band energy diagram incorporating polarization and channel-injected Q\textsubscript{it,ch} contributions in the analytical model. (i-l) Charge-trap flash assisted by ferroelectric polarization. Hybrid gate stack integrating channel-injected charge trapping and field-induced antiferroelectric switching. Polarization-enhanced electric field across the tunnel oxide promotes channel-side charge injection, producing HVT and LVT states governed by charge trapping. Flat-band diagram and analytical model accounting for trapped charge effects. (m-p) Ferroelectric transistor assisted by charge trapping. Hybrid gate stack combining polarization switching with gate-side charge trapping. Synergistic alignment of gate-injected Q\textsubscript{it,gate} and switched polarization enhances the electric field across gate-blocking layer, producing LVT and HVT states governed by coupled mechanisms. Flat-band energy diagram incorporating P\textsubscript{r}, Q\textsubscript{it,gate} and Q\textsubscript{it,ch} effects in the analytical model.}}
    \label{fig:fenand}
\end{figurehere}
\noindent the flat-band condition for the PGM state. At this condition, the electrostatic potential induced by the trapped charge determines the degree of channel band bending, from which we can derive an analytical model for the MW using the flat-band (FB) condition\cite{kim2024depth}. According to this model, the MW increases with greater trapped charge (Q\textsubscript{trap}) and decreases with the capacitance of the charge-trap layer (C\textsubscript{CTL}) and blocking oxide (C\textsubscript{BO}). Although charge-trap-based NAND flash offers high process maturity, its large PGM and ERS voltages (typically exceeding ±20 V) exacerbate the intrinsic trade-off between scaling and reliability.

The ferroelectric transistor employs a metal-ferroelectric-interfacial layer (IL)-semiconductor (MFIS) structure in which information is stored solely through polarization switching (Fig. \ref{fig:fenand}e). During PGM and ERS, polarization reversal modulates the threshold voltage (V\textsubscript{TH}), while partial switching enables multi-level operation (Fig. \ref{fig:fenand}f). A key distinction from charge-trap flash is that the ferroelectric transistor exhibits an LVT state after applying a positive gate voltage (corresponding to PGM) and an HVT state after applying a negative gate voltage (corresponding to ERS) as shown in Fig. \ref{fig:fenand}g. Fig. \ref{fig:fenand}h presents the corresponding band diagram under the flat-band condition, which forms the basis for the analytical model of the MW. The MW scales with the magnitude of polarization switching and inversely with the linear capacitance of the ferroelectric layer (C\textsubscript{FE}), implying that a wider MW inevitably requires a thicker ferroelectric film, which presents a fundamental constraint for device scaling to achieve flash comparable window for high logic density. Furthermore, during PGM/ERS cycling, channel-injected carriers generate interface-trapped charges (Q\textsubscript{it,ch}) at the ferroelectric/IL interface, narrowing the MW and causing read-after-write delay (RAWD), thus emphasizing the need for effective interface control\cite{hoffmann2022fast, wang2021standby, kim2024low}. Because polarization reversal occurs at much lower voltages and shorter timescales than charge trapping, the ferroelectric transistor inherently supports low-power and high-speed operation. Yet its limited MW, channel percolation effects, and device-to-device variability arising from the polycrystalline and polymorphic nature of hafnia ferroelectrics remain key challenges\cite{kim2021cmos, florent2017first, ni2021channel, xiang2020implication, kim2025middle}. To address these critical limitations, especially the narrow MW, a new approach has been proposed. Recently, AlN-based ferroelectrics have gained significant attention as a promising approach to overcome the limited MW achievable with typical Ferroelectric transistors. Wurtzite ferroelectrics possess intrinsically high coercive fields and relatively low dielectric permittivity, offering the potential to simultaneously improve MW and retention when incorporated into the gate stack \cite{kim2023scalable, liu2021post, gao2024ferroelectric}. These materials have enabled a wide range of experimental demonstrations, from proximity-induced switching in AlN-HfZrO heterostructures to reliable multi-level operation, collectively suggesting a viable direction for addressing the fundamental limitations of hafnia-only ferroelectric transistors\cite{skidmore2025proximity, kim2025proximity, kim2025decoupling}. However, translating these advances into practical NAND memory will require further materials and process development to convert sputtered AlN films into ALD processes suitable for 3D NAND integration, as well as device-level engineering to optimize gate configurations\cite{tanim2025atomic}.

The first type of hybrid FeNAND, referred to as charge-trap flash assisted by polarization switching, incorporates an antiferroelectric layer as the blocking oxide (Fig. \ref{fig:fenand}i). Similar to the standard charge-trap flash, data are stored as charges injected from the channel into the Si\textsubscript{3}N\textsubscript{4} charge-trap layer. When a conventional ferroelectric layer is used, however, the remnant polarization (P\textsubscript{r}) has a polarity opposite to that of the trapped charge, causing electrostatic interference that degrades device performance. To mitigate this effect, an antiferroelectric blocking oxide is introduced\cite{kim2023opportunity, ali2020novel}. The antiferroelectric exhibits zero P\textsubscript{r} below its coercive field and undergoes field-induced polarization switching only beyond that threshold, enabling polarization modulation. During PGM, the antiferroelectric layer experiences field-induced polarization reversal, which transiently reinforces the electric field across the tunnel oxide and enhances FN tunneling (Fig. \ref{fig:fenand}j). This effect promotes electron injection from the channel, resulting in improved programming efficiency, expanded MW, and reduced PGM voltage. During read and pass operations, however, both tunneling and field-induced switching must remain suppressed through precise gate-stack engineering to avoid unwanted disturbance. In this hybrid configuration, polarization switching and charge trapping act cooperatively to achieve efficient PGM operation, while charge trapping remains the primary storage mechanism and polarization plays a supportive role. Consequently, the device exhibits an HVT state after PGM and an LVT state after ERS (Fig. \ref{fig:fenand}k). Based on the flat-band condition, the analytical model of MW assumes negligible P\textsubscript{r}, indicating that MW increases with Q\textsubscript{trap} and decreases with C\textsubscript{CTL}, C\textsubscript{BO}, and the capacitance of the antiferroelectric layer (C\textsubscript{AFE}) (Fig. \ref{fig:fenand}l). This analysis reveals that while field-induced polarization enhances programming efficiency, it does not directly contribute to MW formation under READ conditions.

Another type of hybrid FeNAND, referred to as a ferroelectric transistor assisted by charge trapping, introduces a gate-blocking layer between the gate metal and the ferroelectric layer to simultaneously exploit polarization switching and charge trapping mechanisms (Fig. \ref{fig:fenand}m). To align the polarities of ferroelectric polarization and trapped charge, gate-side charge injection is utilized\cite{yoo2024highly, joh2024oxide, kuk2024superior}. During PGM, holes injected from the gate traverse the gate-blocking layer and form interface-trapped charges (Q\textsubscript{it,gate}) at the interface between the blocking oxide and the ferroelectric layer. Or the positive charge in the gate side could also result from electron detrapping from the donor-like traps to the gate metal. The exact mechanisms remain to be investigated. Concurrent with the trapping of positive charge, polarization reversal occurs, and the generated Q\textsubscript{it,gate} not only contributes to MW formation but also compensates the bound charge of the switched polarization, thereby stabilizing the polarization state. The strengthened polarization, in turn, increases the band bending of the gate-blocking layer, promoting additional gate-side charge injection. This mutual reinforcement establishes a positive feedback loop that enables a larger MW at lower operating voltages compared with conventional charge-trap flash (Fig. \ref{fig:fenand}n). In this configuration, both P\textsubscript{r} and Q\textsubscript{it,gate} share the same polarity and cooperatively expand the MW, breaking the typical trade-off between MW and ferroelectric thickness observed in conventional ferroelectric transistors. As a result, the device exhibits an LVT state after PGM and an HVT state after ERS (Fig. \ref{fig:fenand}o). The analytical model derived from the flat-band condition for the programmed state indicates that the MW scales proportionally with Q\textsubscript{it,gate} and P\textsubscript{r}, and inversely with the capacitance of the gate-blocking layer (C\textsubscript{GBL}) and C\textsubscript{FE} (Fig. \ref{fig:fenand}p). However, as in standard ferroelectric transistors, channel-injected Q\textsubscript{it,ch} can still reduce the MW and induce RAWD, emphasizing the need for refined operation schemes and interface engineering to suppress their impact\cite{ma2025investigating, kim2024optimizing, myeong2024comprehensive}.

\begin{figurehere}
   \centering
    \includegraphics[scale=0.1,width=\textwidth]{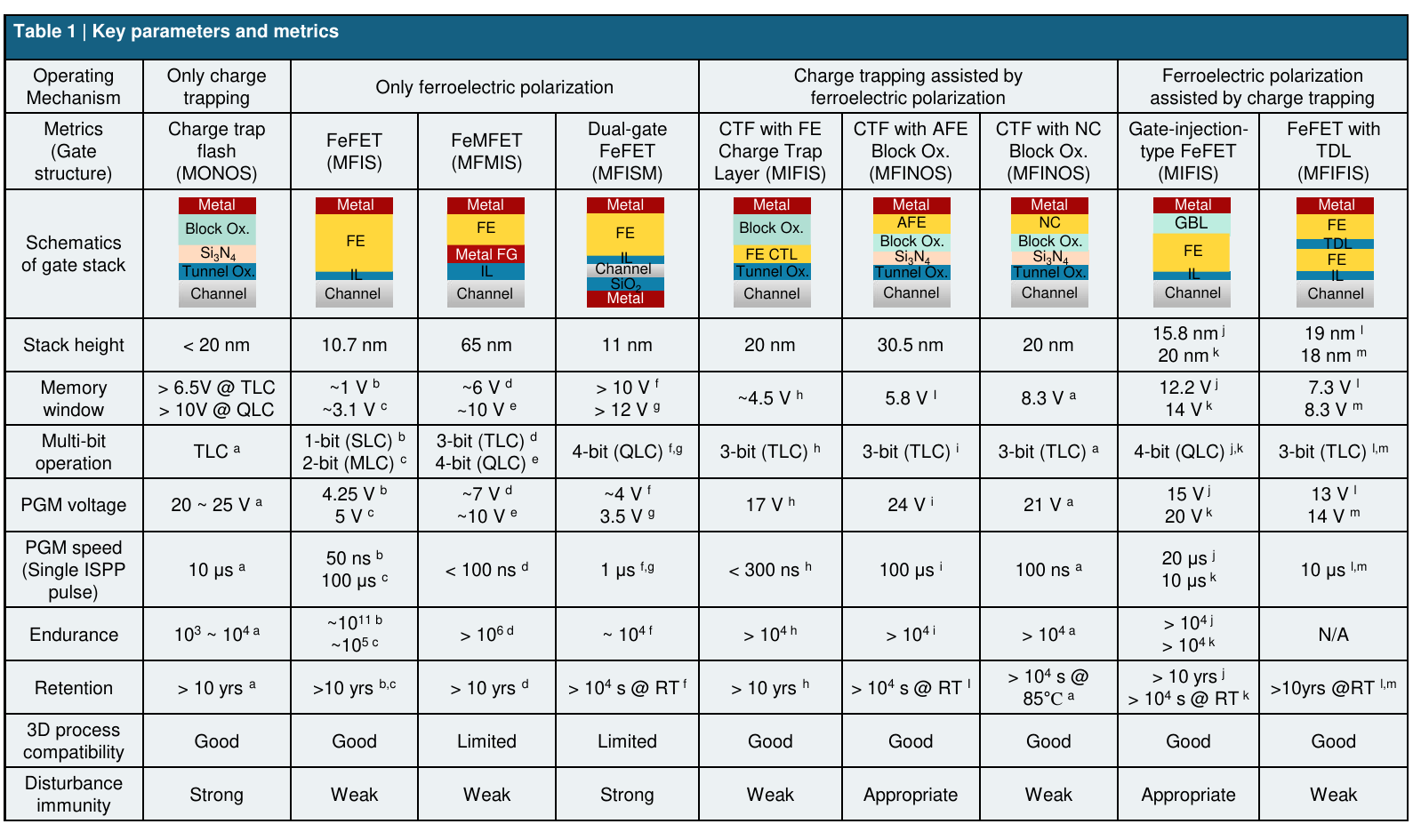}
    \captionsetup{parbox=none} 
    \caption{\textit{\textbf{Hybrid ferroelectric–charge trapping memories.} Key device parameters and performance metrics comparing various hybrid ferroelectric-charge trapping memories. a: ref \cite{kim2023opportunity}, b: ref \cite{zhou20253d}, c: ref \cite{chan2020fefet}, d: ref \cite{kim2022high}, e: ref \cite{kim2024unlocking}, f: ref \cite{mulaosmanovic2021ferroelectric}, g: ref \cite{jiang2022asymmetric}, h: ref \cite{ali2020novel}, i: ref \cite{shin2020capacitance}, j: ref \cite{kuk2024superior}, k: ref \cite{joh2024oxide}, l: ref \cite{das2023experimental}, m: ref \cite{fernandes2024material}, FeFET, ferroelectric field-effect transistor; FeMFET, ferroelectric transistor with floating metal gate; IL, interfacial layer; GBL, gate-blocking layer; TDL, tunnel dielectric layer;}}
    \label{fig:comparison}
\end{figurehere}

Fig. \ref{fig:comparison} compares representative transistor memory architectures that rely on charge trapping, ferroelectric polarization, and their coupled mechanisms. The comparison highlights the intrinsic trade-offs among performance, reliability, and 3D process compatibility, serving as a framework for assessing technological readiness for high-density integration. Conventional charge-trap flash, due to its low charge injection efficiency, offers excellent retention and strong disturbance immunity but requires high PGM voltages above 20 V and endures only about 10\textsuperscript{3} to 10\textsuperscript{4} cycles, limiting further scaling and energy efficiency\cite{kim2023opportunity, joh2024oxide, kuk2024superior, sharma2025wide}. In contrast, ferroelectric field-effect transistors (FeFETs), with its superior polarization switching efficiency, operate at low voltages below 5 V and switch within tens of nanoseconds while maintaining strong intrinsic reliability\cite{feng2024first, zhou20253d, chan2020fefet}. Yet their weak disturbance immunity, a side effect of the high switching efficiency, renders them unsuitable for 3D NAND arrays, in which frequent PASS operations can perturb stored states. Also the fundamental tradeoff between the ferroelectric thickness and memory window also make it not applicable for 3D NAND. Ferroelectric transistor with floating metal gate (FeMFETs) and dual-gate FeFETs deliver the performance and reliability demanded for NAND operation but face structural challenges in vertical scaling. FeMFETs require precise gate-stack engineering to balance the capacitance between the ferroelectric layer and the MOSFET channel, complicating fabrication and reducing process compatibility\cite{kim2022high, ni2018soc}. Moreover, the incorporation of a floating gate further increases the stack thickness beyond 50 nm, lowering integration density\cite{kim2024unlocking}. Dual-gate FeFETs, on the other hand, decouple read and write through independent gates; electrostatic coupling via a thick dielectric read gate can amplify the MW, but in 3D NAND configurations the large equivalent capacitance of the back gate suppresses this effect, limiting practical adoption\cite{mulaosmanovic2021ferroelectric, jiang2022asymmetric, zhao2024paving}. In addition, the thick back gate dielectric not only increases the MW, but also amplifies the front gate variability, leaving the effective number of distinguished bits sensed at the back gate the same as the small MW resulted from front gate operation\cite{chatterjee2022comprehensive}.

Hybrid memories coupling ferroelectric polarization and charge trapping present a promising route for 3D NAND applications. Ferroelectric-assisted charge-trap structures offer stable retention and a moderate MW but still demand further optimization to lower PGM voltage and enhance disturbance immunity\cite{kim2023opportunity, shin2020capacitance, ali2020novel}. In contrast, ferroelectric transistors assisted by charge trapping combine low-voltage operation, fast switching, and strong disturbance robustness with high compatibility for 3D fabrication, positioning them as leading candidates for next-generation NAND integration\cite{fernandes2024material, joh2024oxide, kuk2024superior, das2023experimental, das2024ferroelectric, venkatesan2025pushing}. These devices can be categorized into gate-injection type and tunnel dielectric layer (TDL)-assisted type. In the former, gate-injected electrons are trapped and act synergistically with ferroelectric polarization\cite{fernandes2024material, joh2024oxide, kuk2024superior}. In the latter, channel-injected charges induce flipping across the TDL during polarization switching, building up additional electrostatic potential\cite{ das2023experimental, das2024ferroelectric}. Specifically, the inserted TDL suppresses the de-trapping of charges involved in MW formation and alleviates migration caused by depolarization, thereby enhancing reliability \cite{venkatesan2025pushing,  venkatesan2024demonstration}. Both device types exhibit excellent memory performance, replacing conventional NAND flash architectures. Nonetheless, their reliability remains limited by the intricate interaction between trapped charges and ferroelectric switching, calling for deeper physical understanding and material-device co-optimization. It is important to note that the FeNAND gate-stack specifications, including the parameters in Table 1, are primarily benchmarked under the slow operation (ISPP) demands of high-density NAND flash. However, the intrinsic speed of the ferroelectric layer suggests a strong potential for exploring faster operation regimes, which is relevant for emerging memory applications. Overall, these comparisons delineate a broader transition from charge-trap-based to hybrid ferroelectric memories, in which the cooperative coupling of charge and polarization enables concurrent advances in scalability, speed, and energy efficiency.

\section*{\textcolor{sared}{\large Challenges and Prospects of Gate-injection-type FeNAND}}

Hybrid ferroelectric-charge trapping memories have been explored with various gate-stack configurations, each exhibiting distinct advantages and limitations that could lead to diverse application opportunities. Among them, we focus on the gate-side injection ferroelectric transistor, which introduces new physical mechanisms and holds the greatest promise for future NAND technology. The gate-side injection FeNAND aligns ferroelectric polarization and gate-injected charges, harnessing their cooperative dynamics to overcome the vertical scaling limits of conventional charge-trap 3D NAND. These two mechanisms not only operate independently but also reinforce each other through a positive feedback process that enables low-voltage programming and enhanced scalability. Despite these advantages, several technological challenges remain for practical implementation. Fig. \ref{fig:challenges} revisits these underexplored issues from the viewpoints of architecture, reliability, and operation, outlining potential design strategies and mitigation pathways. Specifically, four key aspects of gate-side-injection FeNAND design are considered: (i) optimization of 3D architecture considering the coupling between ferroelectric polarization and charge trapping, (ii) retention improvement through controlled interaction of the two mechanisms, (iii) enhanced endurance under PGM/ERS cycling, and (iv) mitigation of array-level operation mismatch caused by the opposite V\textsubscript{TH} polarity of PGM/ERS states (LVT/HVT in FeNAND is the opposite to that in conventional flash). In addition, in future 3D NAND technologies, the channel material will be as critical a design component as the gate stack itself\cite{han2023fundamental}. Polycrystalline silicon (poly-Si) channels suffer from low carrier mobility and high thermal budgets, whereas oxide semiconductors (OSs) are emerging as promising channel alternatives\cite{yoo2024highly, joh2024oxide}. Nevertheless, OS-based FeNANDs still face challenges such as limited on-state current and a narrow MW\cite{kim2021cmos, kim2024exploring}. Fig. \ref{fig:challenges} examines these limitations and outlines design strategies for improving performance in OS channel-based FeNANDs.

Gate-side injection type FeNAND exhibits a charge injection path opposite to that of conventional charge-trap flash. Therefore, to maximize the performance of FeNAND and minimize secondary effects, a fundamental rethinking of its 3D architecture is required. In the GAA geometry, the gate-blocking layer lies on the outer surface of the cylinder, while the IL is located inside, creating an electric-field imbalance: the gate-blocking layer experiences a weaker field, whereas the IL is exposed to excessive field concentration\cite{barik2023evolution}. This asymmetry suppresses gate-side charge injection and promotes unwanted channel-side injection, disrupting the cooperative dynamics between the dual mechanisms and ultimately degrading device performance. An architectural paradigm that reflects the intrinsic operating physics of FeNAND is therefore imperative. A more suitable configuration is the channel-all-around (CAA) structure, in which the gate-blocking layer is positioned inside and the IL outside. This geometry enhances the electric field at the gate side while alleviating the field across the IL, enabling efficient charge injection, lowering the PGM voltage, and improving both the MW and reliability. However, as the CAA represents a radical architectural change, a comprehensive evaluation of its potential gains and associated costs is required.

In nonvolatile memories, retention characteristics determine data integrity and directly influence system reliability. Optimizing the retention behavior of gate-side injection FeNANDs has thus become a major focus for further development. Retention degradation in FeNANDs can be classified into short-term and long-term components. The short-term degradation arises from the RAWD effect, primarily caused by channel-injected Q\textsubscript{it,ch} during PGM, which leads to increased system latency and reduced throughput. These injected charges include two distinct populations: those electrostatically compensated by ferroelectric polarization and those remaining uncompensated\cite{10689519, 9720516, 9265055}. Immediately after PGM, the unbalanced charges temporarily elevate the V\textsubscript{th,PGM} above its intrinsic LVT, leading to an under-PGM condition. Over time, these charges gradually detrap, relaxing V\textsubscript{th,PGM} toward equilibrium, and the resulting transient shift is one of the most persistent reliability concerns in FeNAND operation. To mitigate the RAWD issue, optimization of the ferroelectric/IL interfaces is essential. OS-channel-based FeNAND can drastically reduce the RAWD effect by suppressing the formation of unwanted IL layers\cite{chen2022first}. In addition, a detrap pulse can serve as an operation-level solution. A post-PGM detrap pulse effectively suppresses RAWD by selectively removing uncompensated Q\textsubscript{it,ch} while preserving gate-injected Q\textsubscript{it,gate} and switched polarization. The detrap pulse parameters must be carefully optimized to ensure selective charge removal without disturbing the stored states or degrading reliability.

\begin{figurehere}
   \centering
    \includegraphics[scale=0.1,width=\textwidth]{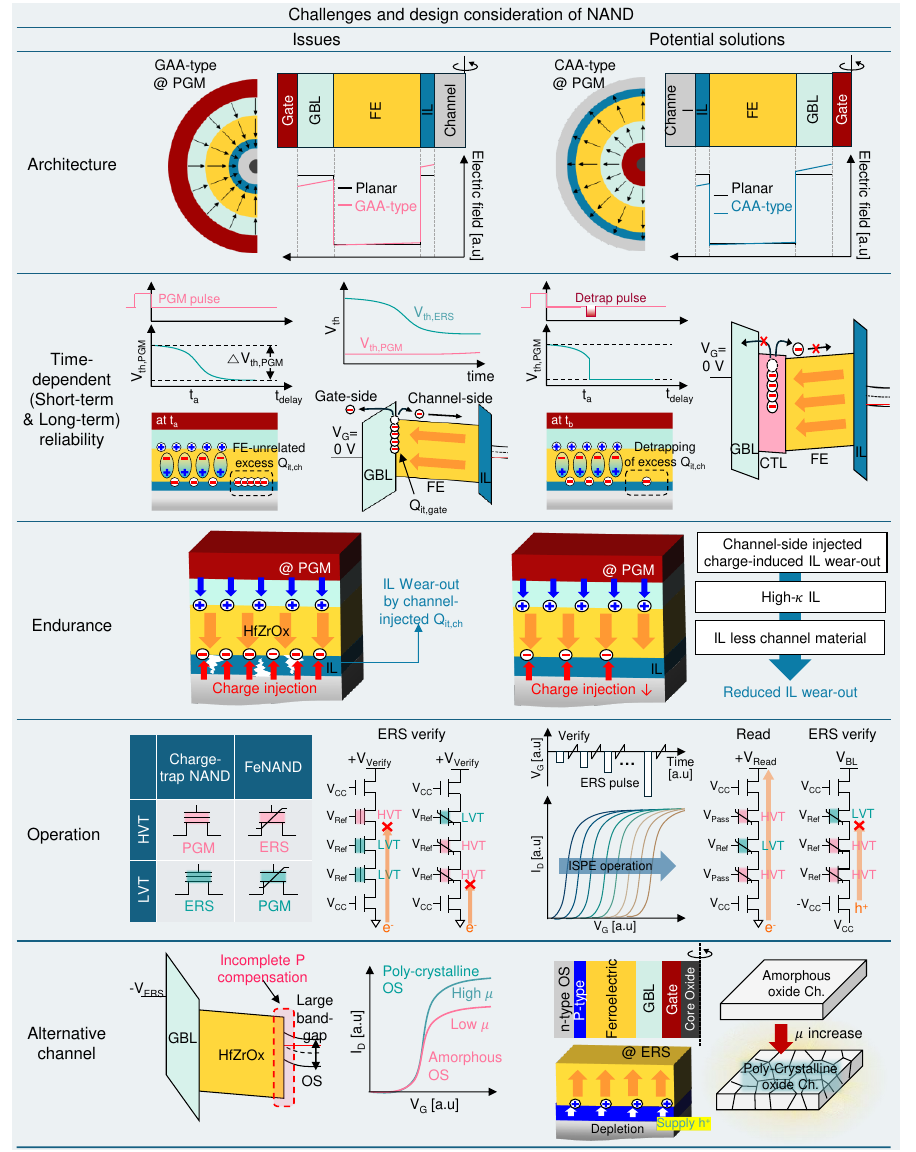}
    \captionsetup{parbox=none} 
    \caption{\textit{\textbf{Current challenges and prospective research directions for FeNAND.} Present limitations of gate-side-injection-type FeNAND memories, summarized in terms of 3D architecture, retention, endurance, array operation and channel materials. Gate-side-injection FeNANDs suffer from field localization, trapped-charge instability and IL degradation, resulting in performance levels below desired device metrics. Future research should emphasize 3D structural optimization, gate-stack and IL engineering, refinement of operation schemes, and OS channel optimization toward commercially viable FeNAND technology.}}
    \label{fig:challenges}
\end{figurehere} 

Long-term retention degradation in FeNANDs predominantly occurs in the ERS state and arises from the detrapping and gradual loss of gate-injected electrons toward either the gate or the channel\cite{kim2025middle, 11075182}. Although the exact physical mechanism of retention degradation remains to be fully understood, it can be aggravated by the positive feedback between charge trapping and polarization switching. Charge detrapping destabilizes the polarization compensation, while depolarization in turn accelerates charge detrapping, forming a self-reinforcing loop. This interplay implies that the coupling between charge trapping and polarization switching can act as both a blessing and a curse. A more strategic approach to improving retention involves engineering the gate stack. One viable direction is the redesining of a charge trap layer (CTL). In addition to serving as a charge storage medium, a properly designed CTL can enhance the P\textsubscript{r} characteristics of the adjacent ferroelectric layer, thereby contributing to improved retention stability\cite{kuk2024superior}. Alternatively, inserting an interlayer between ferroelectric layers can suppress charge detrapping toward the channel and simultaneously enhance P\textsubscript{r}, leading to improved long-term retention\cite{kim2025middle, das2023experimental
, das2024ferroelectric, venkatesan2025pushing}. Such gate-structure engineering should not only improve trapped charge stability but also ensure material compatibility with the ferroelectric layer to maintain polarization performance. Furthermore, the addition of extra gate components must be carefully optimized considering operating voltage.

The endurance characteristic, defined as the ability to withstand repeated PGM/ERS cycling, represents a crucial reliability metric for nonvolatile memories. FeNANDs integrating hafnia ferroelectrics offer low-voltage and fast-switching, thereby promising superior endurance compared with conventional charge-trap NANDs. However, gate-side-injection FeNANDs typically exhibit limited endurance of only 10\textsuperscript{3}-10\textsuperscript{4} cycles, primarily due to the degradation of the gate-blocking layer and IL caused by the repeated generation and detrapping of charges injected from both the gate side and the channel side during PGM/ERS cycling\cite{lim2023comprehensive, yoo2024highly, joh2024oxide, sharma2025wide}. Among these, IL degradation is dominant because of its thin thickness, and suppressing IL wear is therefore a key challenge for endurance improvement. In this regard, employing high-$\kappa$ dielectrics such as Ta\textsubscript{2}O\textsubscript{5} or TiO\textsubscript{2} as the IL can effectively lower field stress during PGM/ERS, enhancing overall device stability\cite{kim2024gate}. Another promising approach is an IL-less configuration using OS-channels. In poly-Si channels, oxygen atoms diffuse from the ferroelectric layer into the channel during high-temperature processing\cite{choi2025unveiling, hwang2025enhanced}. This interdiffusion leads to the formation of an interfacial SiO\textsubscript{2} layer and simultaneously generates oxygen vacancies (V\textsubscript{O}) within the ferroelectric film, both of which degrade polarization stability and accelerate endurance loss\cite{toprasertpong2022breakdown}. In contrast, OS-channels possess chemically stable, oxygen-saturated surfaces that effectively suppress interfacial reactions and prevent interfacial layer formation at the ferroelectric boundary, offering an intrinsically robust platform for long-term cycling. 

The key distinction between gate-side injection type FeNANDs and conventional charge-trap-based NANDs lies in the opposite V\textsubscript{TH} polarity of their PGM and ERS states. This inversion necessitates a fundamental reconfiguration of the FeNAND array operation scheme, since existing protocols were optimized for charge-trap devices. Among the array-level challenges, the ERS-verify operation is particularly critical. In charge-trap NANDs, the ERS state corresponds to a LVT condition. After block-level erase, all cells are expected to be reset to a LVT state. Since NAND cells are connected in series within a string, applying a read voltage to all gates allows current to flow through the channel. If any cell fails to erase and remains at a HVT state (highlighted as pink), that cell blocks the current path, preventing conduction through the string. This current blocking enables clear identification of the defective cell that failed to erase. This process relies on setting V\textsubscript{ref} between the LVT and HVT levels, allowing only erased cells to conduct. In FeNANDs, however, the ERS state corresponds to a HVT condition, rendering the conventional ERS-verify approach inapplicable. Under identical bias conditions, fully erased HVT cells (highlighted in pink) block current through the string, thereby concealing the presence of ERS-failed cells and complicating fault detection during array operation.

One possible approach to overcome this limitation is the incremental-step-pulse-erasing (ISPE) technique, which programs the whole block first, rather than erase the block, and then applies gradually increasing negative voltages to the WL to finely tune V\textsubscript{th,ERS}\cite{park2024reliability, park2022retention}. However, ISPE introduces peripheral-circuit overhead because it requires negative biasing of the WL. In addition, array operation during ISPE, considering disturbance and erase inhibition, still remains an open challenge. Therefore, a new ERS-verify methodology tailored to the operating physics of FeNAND is required. The solutions are inspired by the observation that when an identical V\textsubscript{ref} is applied to all WLs, electrons cannot pass through HVT cells but can traverse LVT cells, while holes exhibit the opposite behavior. Leveraging this property, an ERS-verify scheme utilizing hole conduction can be adopted such that normal operations are performed with electron condution while the verify are done with hole conduction\cite{patent}. In this scheme, band-to-band tunneling between the ground-select line and the common source line injects holes into the channel\cite{kim2023process, ryu2023selective, lim2023improvement}. Erased HVT cells then allow hole transport, whereas ERS-failed LVT cells block it. By monitoring the resulting voltage variation at the bit line (V\textsubscript{BL}), the ERS state of each string can be accurately verified in FeNAND arrays.

An important direction for gate-side injection FeNANDs is identification and optimization of channel materials that are both structurally and functionally compatible with hafnia ferroelectrics. The poly-Si channel commonly used in conventional NAND forms an interfacial layer with the ferroelectric layer, while grain-boundary defects further restrict on-state current, posing a bottleneck for vertical integration. In this context, OS channels have attracted significant attention as promising alternatives for FeNAND applications\cite{kim2021cmos, kim2024exploring, kim2023highly, yoo2024highly, joh2024oxide, choi2025unveiling, hwang2025enhanced}. However, OS channels still require further optimization. Due to their wide bandgap and low hole mobility, OS materials provide insufficient hole density to compensate ferroelectric polarization during ERS operation\cite{yoo2024highly, joh2024oxide}. A practical approach is to insert a hole-supply layer between the OS channel and the ferroelectric layer\cite{choi2025unveiling, hwang2025enhanced, kim2022design, park2025interfacial}. Another important research direction lies in improving the carrier mobility of OS channels. Amorphous OS channels typically exhibit low mobility ($\leq$50 cm\textsuperscript{2}/Vs), which reduces the on-state current and narrows the sensing margin in NAND arrays\cite{conley2010instabilities}. Increasing the grain size and controlling the crystallographic orientation in polycrystalline OS films can effectively reduce grain-boundary scattering and defect-related barriers, enabling high carrier mobility exceeding 100 cm\textsuperscript{2}/Vs\cite{choi2025unveiling, hwang2025enhanced}. Such improvements in channel transport are crucial for maintaining read accuracy and high-speed operation in large-scale FeNAND arrays. Ultimately, comprehensive material-device co-optimization between OS channels and FeNAND architectures will be critical for realizing next-generation ferroelectric memories.

\section*{\textcolor{sared}{\large Opportunities of OS-channel based FeNAND}}
This section explores the synergistic potential of OS channels and hybrid FeNAND for next-generation 3D NAND commercialization, highlighting both performance and process perspectives (Fig. \ref{fig:osc}). Though it is nascent to integrate OS channels with gate-side injection type FeFET, early results show great promise\cite{fernandes2025comparative}. It therefore calls for an urgent need to understand how the OS channels interact with the polarization switching and gate-side charge injection. A deeper understanding of these mechanisms provides critical insight into how such coupling can be exploited to enhance array-level efficiency. From a performance viewpoint, the key advantage of OS-channel FeNAND lies in its ability to markedly reduce array-level power consumption during unavoidable PASS operations. Unlike conventional charge-trap NAND, where the ERS state resides in the negative V\textsubscript{TH} region and multiple PGM states occupy a wide positive range, OS-channel FeNAND exhibits an inverted V\textsubscript{TH} distribution, in which most PGM states lie in the negative region and the ERS state is located near 0 V\cite{yoo2024highly, joh2024oxide}. This polarity shift allows the PASS voltage (V\textsubscript{PASS}) during normal operations to approach 0 V, substantially lowering overall power consumption and eliminating the pass-disturb issue. As vertical stacking progresses, the extended channel length and increased parasitic resistance degrade the on-state current and sensing margin. OS channels offer an effective remedy. While amorphous OS materials provide limited mobility unsuitable for deep-stack arrays, polycrystalline OS channels can achieve high mobility through grain enlargement and orientation control, ensuring sufficient sensing current and high-speed READ operation\cite{han2023fundamental, choi2025unveiling, hwang2025enhanced}. In addition, OS channels form chemically stable, oxygen-saturated interfaces with hafnia ferroelectrics, effectively suppressing interfacial layer formation and excess V\textsubscript{O} generation\cite{toprasertpong2022breakdown}. From a process perspective, OS channels offer clear advantages for both vertical and lateral scaling of FeNAND architectures. In Poly-Si channels, crystallization requires temperatures exceeding 950 $^{\circ}$C, imposing a severe thermal budget incompatible with hafnia ferroelectrics\cite{choi2025unveiling, hwang2025enhanced, 10019458}. By contrast, OS channels can be crystallized at substantially lower temperatures, ensuring excellent process compatibility with ferroelectric layers\cite{kim2023demonstration, rabbi2022polycrystalline}. Moreover, unlike poly-Si, whose crystallinity and carrier transport degrade below 3 nm, OS channels can maintain electrical functionality even at sub-nanometer thickness, enabling further lateral scaling in 3D NAND\cite{si2022scaled}.

\begin{figurehere}
   \centering
    \includegraphics[scale=0.1,width=\textwidth]{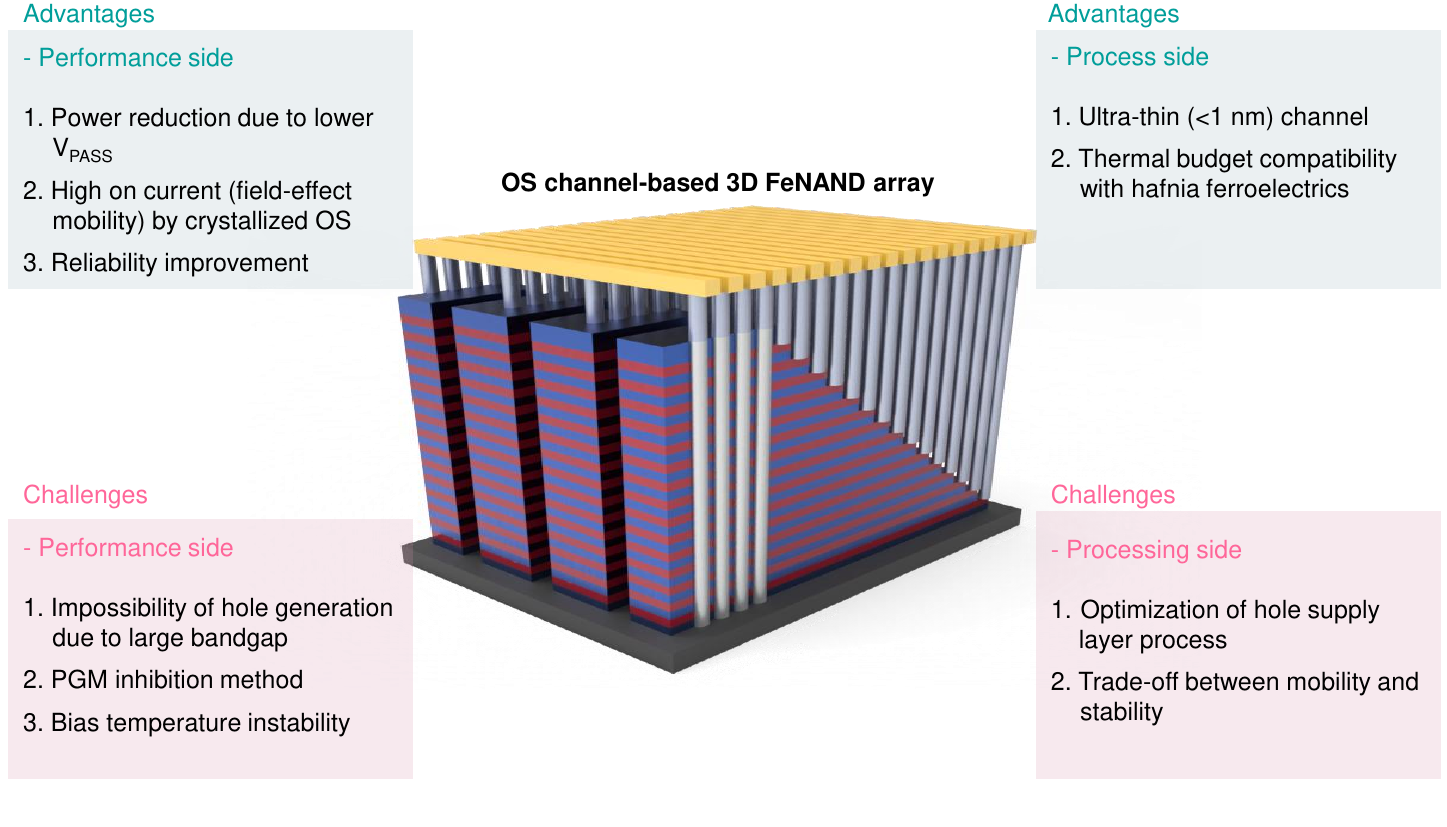}
    \captionsetup{parbox=none} 
    \caption{\textit{\textbf{Opportunities and research outlook for OS-channel based FeNAND.} OS channel-based FeNANDs offer opportunities for performance enhancement and process scalability compared with conventional charge-trap NAND flash. Remaining challenges lie in integrating the synergistic effects of ferroelectric polarization and charge trapping with the intrinsic advantages of OS channels to achieve fully optimized functionality.}}
    \label{fig:osc}
\end{figurehere}

In contrast, OS-channel FeNAND faces unique challenges from both performance and process standpoints. The wide bandgap of the OS channel limits hole generation, hindering complete polarization compensation during erase (ERS) and disabling gate-induced drain-leakage (GIDL)-assisted ERS\cite{kim2023process, ryu2023selective}. Moreover, the opposite \textit{V}\textsubscript{TH} polarities in FENAND necessitate a comprehensive re-examination of NAND array operation. If the conventional incremental step pulse programming (ISPP) scheme is applied to OS-channel FeNAND, three major concerns arise. First, the feasibility of block erase must be investigated. Without holes, achieving uniform raised channel potential during erase requires further scrutiny; otherwise, negative biases may need to be applied to all wordlines (WLs). Second, erase verification becomes even more difficult in OS-channel FeNAND due to the absence of hole conduction, rendering hole-assisted verification unworkable. Both issues could potentially be mitigated by adopting a hybrid channel design that combines an OS channel for electron transport with a Si channel for hole generation, although substantial research is still required to validate this concept. Third, the program inhibition scheme also warrants reconsideration under these altered device physics. In conventional charge-trap NAND, PGM inhibition is achieved through a global or local self-boosting scheme, in which unselected strings are floated and non-selected WLs are biased with a V\textsubscript{PASS} to elevate the channel potential. When the V\textsubscript{PGM} is applied to the target WL, the boosted potential of unselected cells in neighboring strings effectively lowers the gate-to-channel voltage, suppressing unintended disturbance. However, in OS-channel FeNANDs, most PGM states lie in the negative V\textsubscript{TH} region while the ERS state remains near 0 V. As a result, even when a V\textsubscript{PASS} higher than the HVT is applied, unselected cells sharing the same WL remain highly susceptible to PGM disturbance. These characteristics highlight the need for new array-level operation schemes to ensure reliable PGM inhibition in future FeNAND arrays. 

If ISPE is adopted, two critical issues must be addressed. When an entire block is programmed to the LVT state, the verify operation can be readily implemented, similar to that in conventional NAND flash arrays. However, challenges arise in the erase process and erase-inhibition schemes. The lack of hole injection results in significant erase latency; unless the erase speed is improved, adopting ISPE could degrade the overall write throughput of the array. Enhancing the erase speed requires a thorough understanding of the underlying erase mechanisms and further stack optimization. Moreover, selective erase poses an additional challenge, as unselected strings must be effectively inhibited from erasure. Existing program-inhibition schemes developed for NAND flash may need to be revisited and adapted to avoid introducing new reliability concerns.

At last, while polycrystalline OS channels offer high mobility advantageous for NAND scaling, they remain vulnerable to bias-temperature instability (BTI) [90]. This instability primarily originates from V\textsubscript{O} and hydrogen-related defects within the channel or at the gate interface, causing V\textsubscript{TH} shifts under bias stress. The effect becomes more pronounced at elevated temperatures due to accelerated hydrogen diffusion, highlighting the importance of optimizing material composition and interface engineering for improved stability [80]. In addition, OS channels exhibit an inherent trade-off between mobility and reliability: increasing carrier density enhances mobility, yet the oxygen vacancies that supply these carriers also act as trap sites. Achieving long-term stability in OS-based FeNAND devices therefore requires a delicate balance between controlling V\textsubscript{O} concentration and suppressing hydrogen diffusion. Collectively, these insights position OS-channel FeNAND as a transformative candidate for next-generation 3D memories, bridging the gap between material innovation and system-level efficiency. Realizing its full potential will hinge on co-optimizing the ferroelectric stack, channel engineering, and array architecture to achieve both reliability and manufacturability.

\section*{\textcolor{sared}{\large Conclusions}}
The transition from charge-trap to hybrid architectures redefines the foundation of 3D NAND storage memory. This shift introduces a new physical framework in which the energy-efficient polarization switching of hafnia ferroelectrics couples synergistically with the stable charge-storage mechanism of trap layers. Such integration simultaneously advances performance, reliability, and scalability while preserving full CMOS compatibility, positioning FeNAND as a key platform to extend the legacy of flash memory into the data-centric era. The commercialization of FeNAND hinges on the cohesive integration of research efforts across materials, devices, arrays, and architectures. From a materials perspective, enhancing the thermal stability of ferroelectrics, refining interface engineering, and ensuring compatibility with emerging channel materials are essential for achieving both functional integration and scalability in 3D architectures. At the device level, the coupled dynamics of polarization switching and charge trapping should be recognized as new design parameters that redefine reliability and operational stability. At the array level, operation schemes incorporating ferroelectric switching, along with optimized detrap pulse, PGM inhibition and ERS verify algorithms, are required to secure reliable NAND functionality. Finally, at the architectural level, a new 3D design paradigm is needed to systematically exploit the complementary nature of the dual mechanisms, moving beyond bit-density scaling toward a balanced optimization of performance, reliability, and energy efficiency.

\section*{\large Data availability}
This perspective paper does not contain any new experimental data. All data discussed are available from the cited literature.

\bibliography{ref}
\bibliographystyle{naturemag}
\section*{\large Acknowledgments}

This work, conducted at the University of Notre Dame, was primarily supported by Samsung Electronics (IO240919-10902-01) and partially supported by the SUPREME Center, one of the SRC and DARPA JUMP 2.0 centers, and NSF 2344819.

This work, conducted at KAIST, was supported by Samsung Electronics, Co., Ltd (IO241203-11376-01), MOTIE (Ministry of Trade, Industry and Energy), Korea (2410011938, 00235655, 23006-15TC and 2410010173, 00231985, 23005-30FC), Ministry of Science and ICT, Korea (RS-2023-00260527).

\section*{\large Author contributions}

G.K. and H.C. led the project and the manuscript preparation.
S.J. and K.N. proposed the project and reviewed the manuscript.
All authors contributed to the discussions.
G.K. and H.C. contributed equally to this work.
S.J. and K.N. are the corresponding authors.

\section*{\large Competing interests}
The authors declare that they have no competing interests.

\end{document}